\documentclass[%
 reprint,
 superscriptaddress,
 showkeys,
nofootinbib,
 amsmath,amssymb,
 aps,
 prd,
]{revtex4-2}

\usepackage[utf8]{inputenc}
\DeclareUnicodeCharacter{2500}{\textemdash{}}

\usepackage[normalem]{ulem}
\usepackage{xcolor}
\usepackage{cancel}

\usepackage{textcomp}
\usepackage{soul}
\usepackage{float}
\usepackage[caption=false]{subfig} 
\usepackage{gensymb}
\usepackage{graphicx}
\usepackage{dcolumn}
\usepackage{bm}
\usepackage[hypertexnames=true,colorlinks=true,urlcolor=blue,bookmarks=true,citecolor=blue,breaklinks=true]{hyperref}
\usepackage[dvipsnames]{xcolor}
\usepackage{booktabs}
\usepackage{makecell}
\usepackage{array}
\usepackage{multirow}
\usepackage[normalem]{ulem}
\newcolumntype{C}[1]{>{\centering\arraybackslash}p{#1}}

\begin{document}

\preprint{APS/123-QED}

\title{Inferring Properties of Rotating Core-Collapse Supernovae with Gravitational-Wave Detector Networks}

\author{A.~Akhmetali}
\email{almat.akhmetali@nu.edu.kz} 
\affiliation{Energetic Cosmos Laboratory, Nazarbayev University, 010000 Astana, Kazakhstan} 
\affiliation{Department of Physics, Nazarbayev University, 010000 Astana, Kazakhstan}
\affiliation{Department of Electronics and Astrophysics, Al-Farabi Kazakh National University, 050040 Almaty, Kazakhstan} 

\author{Y.~S.~Abylkairov}
\email{sultan.abylkairov@nu.edu.kz}
\affiliation{Energetic Cosmos Laboratory, Nazarbayev University, 010000 Astana, Kazakhstan} 
\affiliation{School of Artificial Intelligence and Data Science, Astana IT University, 010000 Astana, Kazakhstan}

\author{S.~Nunes}
\affiliation{Centro de Física das Universidades do Minho e do Porto (CF-UM-UP),
Universidade do Minho, 4710–057 Braga, Portugal}

\author{J.~A.~Font}
\affiliation{Departamento de Astronomía y Astrofísica, Universitat de València, Avinguda Vicent Andrés Estellés 19, 46100 Burjassot (Valencia), Spain}
\affiliation{Observatori Astronòmic, Universitat de València, Catedrático José Beltrán 2, 46980, Paterna, Spain}

\author{M.~Zanolin}
\affiliation{Embry-Riddle Aeronautical University, 3700 Willow Creek Road, Prescott, Arizona 86301, USA}

\author{E.~Abdikamalov}
\affiliation{Energetic Cosmos Laboratory, Nazarbayev University, 010000 Astana, Kazakhstan}
\affiliation{Department of Physics, Nazarbayev University, 010000 Astana, Kazakhstan}

\date{\today}

\begin{abstract}
Core-collapse supernovae are among the most promising yet still undetected sources of gravitational waves. A future detection would provide a direct view of the physical processes occurring deep inside a collapsing star. In this work, we investigate how networks of current and future gravitational-wave detectors can constrain the properties of rapidly rotating core-collapse supernovae using their characteristic core-bounce and early post-bounce signals. Using deep-learning techniques, we estimate the peak frequency, rotation rate, and signal amplitude from noisy detector data and compare the performance of different detector-network configurations. We find that detector networks improve both parameter recovery and sky coverage. For current-generation networks, estimation of the peak frequency is possible out to about 15 kpc, while the rotation rate remains recoverable out to distances exceeding 100 kpc. Third-generation observatories extend these distances by approximately a factor of three to four.
\end{abstract}


\maketitle

\section{Introduction}
\label{sec:intro}

Gravitational waves (GWs) provide a direct probe of the most extreme environments in the universe. While compact binary mergers are now routinely detected through their GW emission \citep{Abac26GWTC5}, core-collapse supernovae (CCSNe) remain a major predicted but as yet unobserved class of GW sources \citep{Mueller26GWREview}. Once detected, CCSN GWs are expected to reveal unique information about the supernova central engine \citep{mezzacappa25colloquium}.

CCSNe occur when a massive star exhausts its nuclear fuel and its core collapses under gravity. The collapse releases $\sim 10^{53}$ erg of gravitational binding energy, most of which is emitted in neutrinos \citep{Janka16Physics, muller20hydrodynamics}. When the inner core reaches nuclear density, the collapse halts and launches a bounce shock that subsequently stalls because of nuclear dissociation and neutrino losses. Reviving this shock is the central challenge of the CCSN explosion mechanism and remains an active area of research \citep{burrows13colloquium, mezzacappa20physical}. Electromagnetic observations provide valuable information about the explosion outcome but only limited access to the engine itself, since the shock is revived within the first second after core bounce while still buried deep inside the stellar envelope. In contrast, neutrinos and GWs escape directly from the core and carry information about the earliest phases of the explosion.

The multidimensional dynamics of the supernova engine naturally generate GW emission \citep{kotake:13review, gossan16observing, abdikamalov22gravitational}. In neutrino-driven explosions, convection, standing accretion shock instability, and proto-neutron star (PNS) oscillations produce GW signals spanning a wide range of frequencies \citep{Murphy09Model, mueller:13, radice:19gw, TorresForne19Towards, Sotani24Universality, kuroda16, Andresen17Gravitational, Hayama18Circular, Mezzacappa20Gravitational}. Additional low-frequency emission may arise from anisotropic neutrino emission and asymmetric shock propagation \citep{mueller:97, Takiwaki18Anisotropic, Vartanyan20GWanisotropy, Choi24GW}. Detecting these signals would provide a direct view of the explosion dynamics before electromagnetic emission becomes observable \citep{Szczepanczyk21Detecting, Nakamura:2016kkl}.

In rapidly rotating progenitors, which likely constitute a minority of CCSN sources \citep{Heger05Presupernova, woosley:06}, the GW signal differs substantially \citep{Sykes26Trends}. Rotation deforms the collapsing inner core, producing a strong GW burst at core bounce followed by a short ring-down phase associated with quadrupolar oscillations of the newly formed PNS \citep{Dimmelmeier08, ott12correlated, Fuller15SNseismology}. Some models additionally develop post-bounce non-axisymmetric instabilities that generate longer-lasting GW emission \citep{Scheidegger08, Shibagaki20new, Pan21Stellar}. Rapid rotation may also lead to magnetorotational explosions, in which rotational energy is transferred to the shock through magnetic fields and drives bipolar outflows \citep{burrows:07b, moesta:14b, kuroda:20, obergaulinger:20}. Additional GW emission may arise from resonances involving PNS oscillation modes \citep{Cusinato:26} and from jet dynamics \citep{Birnholtz13GW_jet, Soker23GWJJ}.

A detection of CCSN GWs would enable direct inference of the physical properties of the source \citep[e.g.,][]{pajkos21, CasallasLagos23Characterizing}. Rotation leaves particularly strong imprints on the GW signal, allowing one to distinguish between neutrino-driven and magnetorotational explosions \citep{Logue12Inferring, Powell24Determining,Villegas:2023wsu} and to constrain the angular momentum of the collapsing core \citep{abdikamalov:14, pajkos19}. GW observations may also provide information about PNS masses and radii \citep{Bizouard21Inference, Bruel23Inference}, the equation of state (EOS) of dense matter \citep{Rover2009, richers:17, edwards17, chao22determining, Wolfe23GW, mitra24, abylkairov2025assessing, Akhmetali26Toward, Akhmetali26EOS, Murphy24Dependence, Rusakov26Exploration, Powell25Impact}, phase transitions in the PNS interior \citep{abdikamalov:09, zha:20}, and black-hole formation following failed explosions or fallback accretion \citep{cerda:13, Pan18Equation, Shibagaki21, Burrows23Black, Powell25noEMCCSN, Eggenberger25Black, Ott11PRL, Kuroda23Failed}.

In our previous work \citep{Akhmetali26PE}, we investigated parameter estimation from the core-bounce and early ring-down GW signals of rotating CCSNe using machine-learning techniques. This phase of the signal is relatively straightforward to model numerically, allowing us to generate thousands of waveforms at modest computational cost and thereby enabling a machine-learning-based approach \citep{mitra23, Abylkairov24Evaluating, Sakan25Probing}. Building on early works \citep{pastor24, nunes2024deep, Villegas25Parameter}, we examined how physical uncertainties, including variations in the exact bounce time and in the source inclination, affect the inference of the GW peak frequency and amplitude at bounce along with the core rotation across a range of progenitor models and nuclear EOS. We demonstrated that Fourier-domain analysis mitigates bounce-time uncertainty and that next-generation GW detectors can constrain core rotation beyond 100~kpc for favorable source orientations. However, those results were obtained for individual detectors and therefore did not account for the additional information available from a global detector network.

Real GW observations are performed by networks of geographically separated detectors. The relative locations and orientations of these instruments determine the network sensitivity, sky coverage, duty cycle, and ability to reconstruct the two GW polarizations. Consequently, detector-network configuration plays a central role in the inference of source properties. The current global second-generation (2G) network consists of the two Advanced LIGO detectors \citep{Aasi:2014jea}, Advanced Virgo \citep{Acernese15}, and KAGRA \citep{Aso13}, and will be expanded by LIGO India \citep{Saleem22}. Looking further ahead, third-generation (3G) observatories such as the Einstein Telescope (ET) \citep{2026:ET} and Cosmic Explorer (CE) \citep{Reitze19,Evans21} will dramatically improve sensitivity and detection reach \citep{srivastava19detection, Srivastava22}.

Recently, \citet{Bruel23Inference} demonstrated that coordinated observations with 2G and 3G detectors can significantly improve the recovery of PNS properties for non-rotating CCSN models. Using the information encoded in the simulated GW signal associated with specific oscillation modes of the PNS, \citet{Bruel23Inference} showed that, assuming unambiguous mode identification, it may be possible to infer the time evolution of the surface gravity of the PNS, $M_{\rm PNS}/R^2_{\rm PNS}$, up to several hundreds of kpc from Earth. (Here, $M_{\rm PNS}$ and $R_{\rm PNS}$ denote the mass and the radius of the PNS, respectively.) Whether these gains extend to the core-bounce and early ring-down signals of rapidly rotating CCSNe remains unexplored. In this work, we investigate this question by examining how detector-network configurations influence parameter estimation, sky localization, and distance reach for rotating CCSN GW signals. We further assess the improvements expected from future GW observatories. As in our previous work \citep{Akhmetali26PE}, we focus on inferring three key source parameters: the peak frequency, $f_{\mathrm{peak}}$, the distance-normalized peak amplitude, $D\Delta h$, and the rotational parameter at core-bounce, $T/|W|$. These quantities encode the core structure and bounce dynamics. Here, $D\Delta h$ is the difference between the maximum and minimum GW bounce signal strain for an optimally oriented observer, while $f_{\mathrm{peak}}$ corresponds to the dominant high-frequency feature of the postbounce waveform and traces PNS ring-down oscillations during the early post-bounce phase. The rotation parameter $T/|W|$, defined as the ratio of rotational kinetic energy to gravitational binding energy at bounce, determines the centrifugal deformation of the inner core and thus the mass quadrupole moment that drives GW emission \citep{Dimmelmeier08}. For $T/|W| \lesssim 0.06$, the amplitude scales approximately as $D\Delta h \propto T/|W|$. At higher rotation rates, however, centrifugal support slows the collapse and bounce dynamics, offsetting the increased deformation and causing the GW amplitude to saturate rather than continue to grow with increasing rotation. In this regime, Coriolis forces can also excite inertial modes that modify both the signal amplitude and frequency \citep{richers:17}. Because $D\Delta h$ and $f_{\mathrm{peak}}$ probe complementary aspects of the bounce and ring-down dynamics, jointly measuring them helps break degeneracies with $T/|W|$. This motivates treating $D\Delta h$, $f_{\mathrm{peak}}$, and $T/|W|$ as independent inference targets.

The remainder of this work is organized as follows: In Section~\ref{sec:detectors}, we provide an overview of the GW detectors considered in our work. Section~\ref{sec:methods} details the waveform dataset, the signal injections and preprocessing, and the methodology we employ for the inference. Section~\ref{sec:results} presents our results and analysis. Finally, Section~\ref{sec:conclusion} provides a summary of our core results and concluding remarks.

\section{Detector Network}
\label{sec:detectors}

Ground-based GW observatories are broadly classified into 2G and 3G detectors. The current global 2G network consists of the two Advanced LIGO~\citep{Aasi:2014jea} interferometers in Hanford and Livingston (USA), the Advanced Virgo~\citep{Acernese15} detector in Italy, and the KAGRA~\citep{Aso13} detector in Japan. The LIGO detectors have 4-km arm lengths, while Virgo and KAGRA operate with 3-km arms. This network will be further expanded by LIGO Aundha~\citep{Saleem22}, a third LIGO-like interferometer under construction in Aundha, India, expected to become operational around 2030. Operating as a coordinated global network, these observatories conduct joint observing runs interleaved with commissioning and upgrade periods aimed at progressively improving their sensitivity toward design goals~\citep{Abbot20}.

The next generation of GW observatories are expected to substantially extend the observational reach of current detectors~\citep{ET_CE_noise}. The main 3G facilities are the ET in Europe and the CE in the USA. The ET design is considering two concepts: a two-site L-shaped (2L) configuration consisting of two 10 km interferometers separated by approximately 1200 km and a triangular underground observatory. In this work, we adopt the triangular ET-D design, consisting of six 10-km arms arranged in an equilateral triangle with $60^\circ$ opening angles between arms. Each pair of arms forms an interferometer, enabling full-sky coverage and redundant signal reconstruction. The ET-D concept further adopts a xylophone configuration, combining cryogenic low-frequency interferometers with room-temperature high-frequency interferometers to suppress seismic and thermal noise while maintaining excellent high-frequency sensitivity. In contrast, the CE design consists of a network of L-shaped interferometers with arm lengths up to 40 km, with staged implementations including 20-km and 40-km facilities. With significantly longer arm lengths and advanced noise mitigation strategies, CE detectors are expected to achieve an order-of-magnitude improvement in strain sensitivity over current-generation observatories~\citep{ET_CE_noise}.

\begin{figure}[t!]
\centering
\includegraphics[width=1\linewidth]{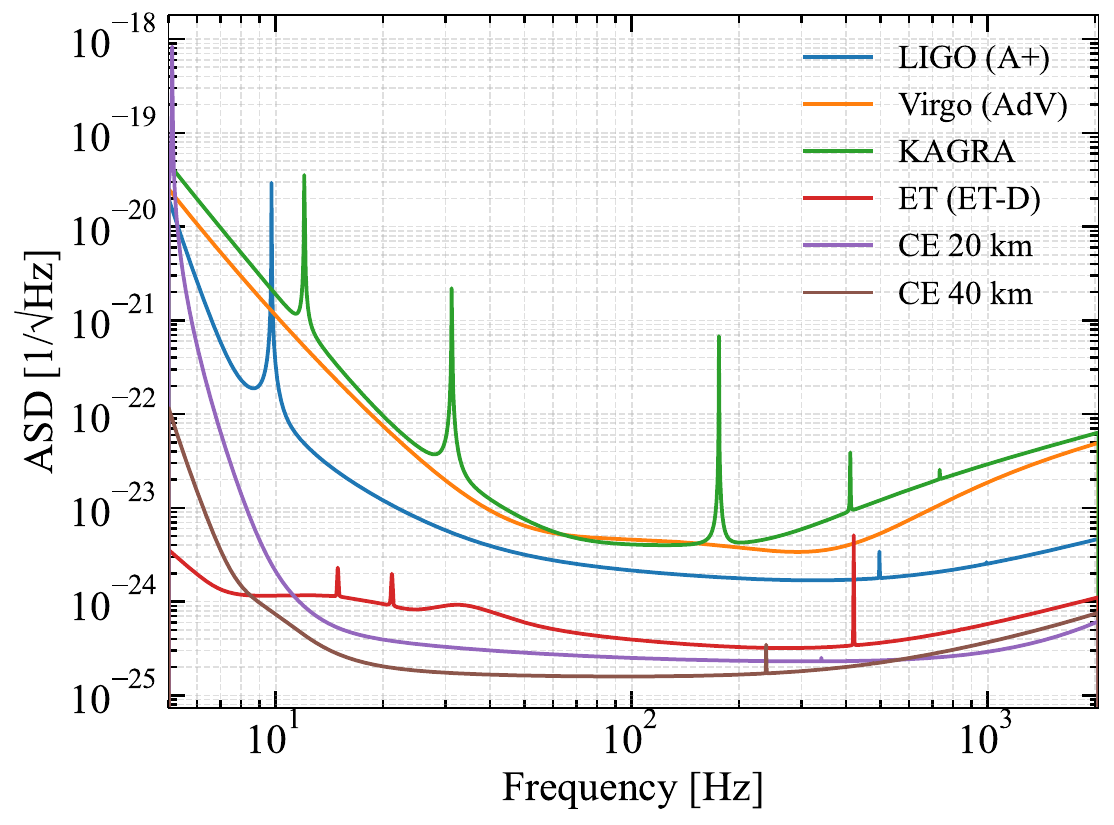}
    \caption{Amplitude spectral densities as a function of frequency for the ground-based detectors considered in this work.} 
     \label{fig:ASDs}
\end{figure}

The response of any of these detectors to an incoming GW depends on the source sky location, polarization, and detector orientation. The measured strain in detector $k$ can be written as~\citep{Schutz11} 
\begin{equation}
h_k(t) = F_{+,k}(\theta,\phi,\psi)\,h_{+}(t) + F_{\times,k}(\theta,\phi,\psi)\,h_{\times}(t),
\label{eq:1}
\end{equation}
where $h_{+}(t)$ and $h_{\times}(t)$ are the two GW polarization modes, and $F_{+,k}$ and $F_{\times,k}$ are the corresponding antenna pattern functions. The angles $\theta$ and $\phi$ define the source position on the sky, while $\psi$ denotes the polarization angle. The antenna pattern functions depend on both the geographic location and the arm orientation of each detector.

The configuration of a detector network in the Earth-fixed frame is fully specified by the coordinates of the beam-splitter and the orientation vectors of the interferometer arms. The corresponding parameters for both 2G and 3G detectors considered in this study are listed in Table~\ref{tab:detectors}. Since the exact locations and orientations of 3G detectors are yet to be determined, following~\citet{Bruel23Inference} we place them at the following locations: the ET is placed at the Virgo site with an arbitrary orientation, while the two CE detectors are located in Idaho and New Mexico (USA)~\citep{Borhanian21}.

\begin{table*}[t]
    \centering
    \resizebox{\textwidth}{!}{
    \begin{tabular}{c|c|c|c|c|c|c|c|c|c|c|c|c}
        \hline
        Detector & $\lambda$ [$^\circ$] & $\phi$ [$^\circ$] & $h$ [m] & $\Psi_1$ [$^\circ$] & $\Psi_2$ [$^\circ$] &
        Detector & $\lambda$ [$^\circ$] & $\phi$ [$^\circ$] & $h$ [m] & $\Psi_1$ [$^\circ$] & $\Psi_2$ [$^\circ$] & $\Psi_{\rm GC}$ [$^\circ$] \\
        \hline
        LIGO Hanford (H)     & -119.41 & 46.46 & 142.55 & 324.00 & 234.00 &
        Einstein Telescope 1 (ET) & 10.50 & 43.63 & 0.0 & 89.95 & 29.96 & -- \\
        
        LIGO Livingston (L)  & -90.77 & 30.56 & -6.57 & 252.28 & 162.28 &
        Einstein Telescope 2 (ET) & 10.63 & 43.63 & 0.0 & 330.04 & 270.05 & -- \\
        
        Virgo (V)            & 10.50 & 43.63 & 51.88 & 19.43 & 289.43 &
        Einstein Telescope 3 (ET) & 10.57 & 43.71 & 0.0 & 210.00 & 150.00 & -- \\
        
        KAGRA (K)            & 137.31 & 36.41 & 414.18 & 60.40 & 330.40 &
        Cosmic Explorer 40 km (CE40) & -112.83 & 43.83 & 0.0 & 180.00 & 90.00 & 27.02 \\
        
        LIGO Aundha (A)      & 77.03 & 19.61 & 0.0 & 332.38 & 242.38 &
        Cosmic Explorer 20 km (CE20) & -106.48 & 33.16 & 0.0 & 240.00 & 150.00 & 83.04 \\
        \hline
    \end{tabular}
    }
    \caption{Coordinates of current (left) and future (right) ground-based observatories. The longitudes $\lambda$ and latitudes $\phi$ specify the beam-splitter (corner-station) locations, while $h$ denotes the elevation above the WGS84 reference ellipsoid in meters. $\Psi_1$ and $\Psi_2$ are the arm orientation angles measured clockwise from local North. For the Cosmic Explorer detectors, $\Psi_{\rm GC}$ denotes the orientation of the first interferometer arm relative to the great circle connecting the two CE sites. The parameters for LIGO Hanford, LIGO Livingston, Virgo, KAGRA, and LIGO Aundha are taken from~\citep{lalsuite}. The exact locations and orientations of 3G detectors are not yet determined. For this work, we follow~\cite{Bruel23Inference} and place the ET at the Virgo site with an arbitrary overall orientation. The CE sites follow~\citep{Borhanian21}, with a 40~km detector located in Idaho and a 20~km detector located in New Mexico.}
    \label{tab:detectors}
\end{table*}

In addition to the detector's response functions, the detectors are characterised by their noise properties. For the 2G network, we adopt the projected design sensitivities of Advanced LIGO (A+)~\citep{barsotti2018a+}\footnote{{\tt aLIGOAPlusDesignSensitivityT1800042} from {\tt PyCBC}}, Advanced Virgo~\citep{Abbot20}\footnote{{\tt AdvVirgo} from {\tt PyCBC}}, and KAGRA~\citep{Abbot20}\footnote{{\tt  KAGRADesignSensitivityT1600593} from {\tt PyCBC}}. For the 3G network, we use the ET-D sensitivity curve for the ET~\citep{Hild2011ET}\footnote{{\tt EinsteinTelescopeP1600143} from {\tt PyCBC}} and compact-binary optimized sensitivity curves for CE~\citep{Srivastava22}\footnote{{\tt CosmicExplorerP1600143} for 40 km and {\tt CosmicExplorerPessimisticP1600143} for 20 km from {\tt PyCBC}}. These amplitude spectral densities, shown in Fig.~\ref{fig:ASDs}, are used to generate colored Gaussian noise into which the simulated GW signals are injected.

We identify a network of detectors with the letters corresponding to each of its detectors (e.g. HLVKA is the network composed of LIGO Hanford (H), LIGO Livingston (L), Virgo (V), KAGRA (K) and LIGO Aundha (A)).

\section{Dataset and methodology}
\label{sec:methods}

Below we describe the waveform dataset, the preprocessing steps of the data, and the deep learning (DL) model used to estimate physical parameters of the source from its GW signals.

\subsection{Data}
\label{sec:data}

The waveform catalog used in this work is identical to that of our previous work \citep{Akhmetali26Toward}\footnote{Available at https://zenodo.org/records/17579189}; we refer the reader there for details and summarize only the essentials here. Waveforms are generated with the {\tt CoCoNuT} code \citep{Dimmelmeier02a,dimmelmeier:05MdM} using axisymmetric simulations, which are adequate through collapse, bounce, and the early post-bounce phase \citep{ott:07cqg}. Deleptonization during collapse is modeled with a density-dependent electron fraction prescription, $Y_e(\rho)$ \citep{Liebendoerfer05Simple}, while post-bounce neutrino effects are treated with a leakage/heating scheme \citep{Ott13General}. The catalog spans four solar-metallicity progenitors ($12$, $15$, $27$, and $40\,M_\odot$) \citep{woosley:07,Heger05Presupernova,whw:02}, six EOSs (\texttt{BHB}$\Lambda \Phi$ \cite{bhbeos}, \texttt{SFHo} \cite{steiner:13b}, \texttt{SFHx} \cite{steiner:13b}, \texttt{LS220} \cite{lseos:91}, \texttt{HSDD2} \cite{hempel:10,hempel:12}, \texttt{GShenFSU2.1} \cite{gshen:11b}), and rotation rates in the range $T/|W| \in [0.02,0.21]$, yielding 1332 waveforms sampled at 4096 Hz. We analyze only the $[-2,6]$ ms interval around bounce to isolate the bounce and early ring-down signal. Later times are excluded because prompt convection is not accurately captured in 2D simulations; however, previous work showed that its inclusion has little impact on parameter estimation performance, indicating that the inference is driven primarily by bounce and ring-down features \citep{Akhmetali26Toward,Akhmetali26PE}.

\subsection{CCSN population}
\label{sec:CCSN}

To account for the nonuniform spatial distribution of CCSNe in the Milky Way, we adopt a Galactic population model based on an exponential stellar disk, following the prescription considered by~\citet{adams13observing}. The spatial density is described by
\begin{equation}
\rho(R,z)\propto \exp\left(-\frac{R}{R_d}\right)\exp\left(-\frac{|z|}{h_z}\right),
\label{eq:CCSN_population}
\end{equation}
where $R$ is the Galactocentric cylindrical radius, $z$ is the height above the Galactic plane, $R_d=2.9$~kpc is the radial scale length, and $h_z=0.095$~kpc is the vertical scale height.

We generate 1332 source positions, corresponding to the number of waveforms in our catalogue. The radial positions are sampled from the distribution $p(R)\propto R\exp(-R/R_d)$, while the azimuthal angles are drawn uniformly from $[0,2\pi)$. The vertical positions follow a symmetric exponential distribution. Assuming a Sun--Galactic-center distance of 8.2~kpc, the generated positions are transformed into equatorial coordinates to determine the source right ascension and declination.\footnote{We use Astropy~\citep{Astropy} to transform coordinates.}

\begin{figure*}
    \centering
    \includegraphics[width=\linewidth]{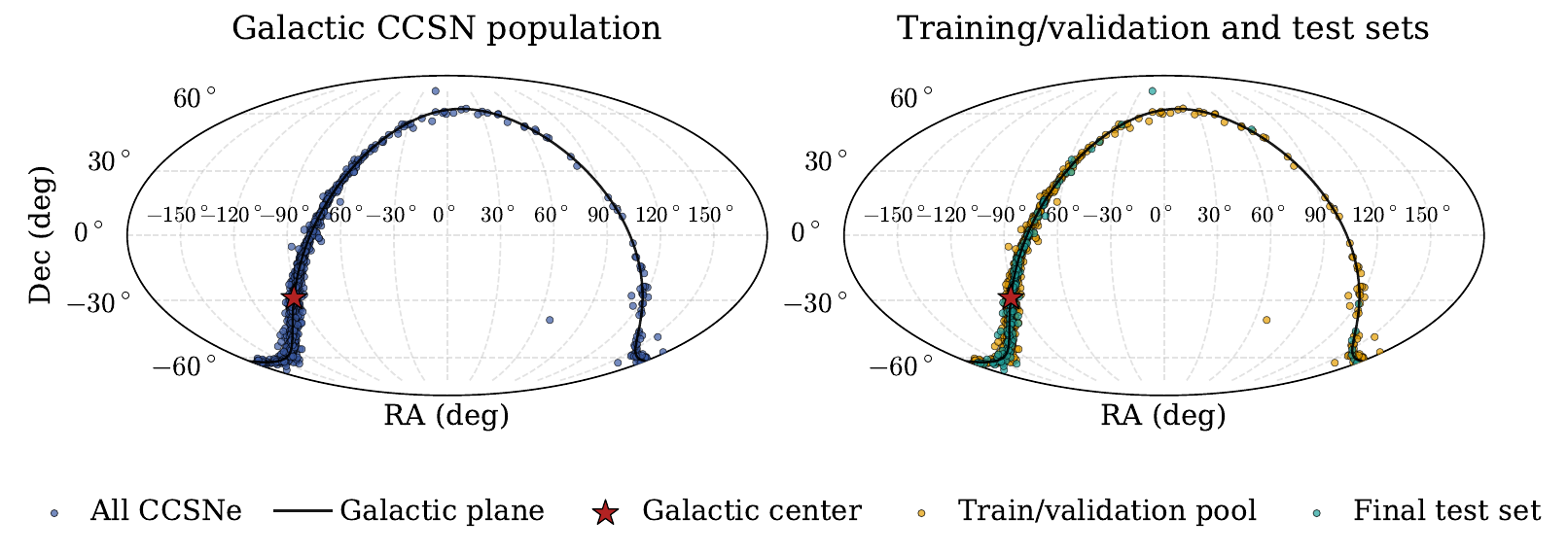}
    \caption{Sky distribution of the 1332 simulated Galactic CCSNe in equatorial coordinates. The left panel shows the complete population (blue points), while the right panel distinguishes the development set (orange points, 85\% of the catalogue), used for training, validation, model comparison, and architecture optimization, from the independent test set (green points, 15\%), reserved exclusively for the final evaluation presented in the Section~\ref{sec:results}. The solid black curve represents the Galactic plane, and the red star marks the Galactic center. The population follows an exponential disk model with a radial scale length of $R_d=2.9$~kpc and a vertical scale height of $h_z=0.095$~kpc~\citep{adams13observing}.}
    \label{fig:CCSN_population}
\end{figure*}

The resulting angular distribution is shown in Fig.~\ref{fig:CCSN_population} (left panel). As expected from the adopted disk model, the simulated sources are strongly concentrated toward the Galactic plane, with a particularly high apparent concentration in the direction of the Galactic center. This distribution provides a more physically motivated representation of the expected Galactic CCSN directions than uniform sampling across the celestial sphere.

Each waveform is assigned a unique sky position, and this assignment remains fixed throughout the analysis. The same population is used for all detector-network configurations and training iterations, ensuring a consistent comparison of their performance.

The sampled Galactic positions are used only to determine the angular distribution of the sources. To investigate the dependence of parameter-estimation performance on detector sensitivity and source distance, all signals are injected at a common prescribed distance within each experiment. Consequently, the individual heliocentric distances obtained from the population model are not used to scale the GW amplitudes. We note that although the source positions are sampled from a Galactic CCSN population, we extend the injection distances well beyond the Milky Way to explore the distance reach of the parameter-estimation framework. At these larger distances, the adopted angular distribution is no longer representative of a realistic extragalactic CCSN population.

\subsection{Signal injection and data preprocessing}
\label{sec:injection}

We generate simulated detector noise using the design sensitivity curves of the detectors discussed in Section~\ref{sec:detectors} with the \texttt{PyCBC} library~\citep{pycbc}, sampled at 4096~Hz. The noise is assumed to be stationary and Gaussian, fully characterized by the detector power spectral density (PSD). 

We consider a network of detectors indexed by $k \in \{1,\dots,N\}$. The response of detector $k$ to a GW signal depends on the antenna pattern functions (Eq.~(\ref{eq:1})) and the geometrical time delay between the detector and the Earth center. The strain measured at detector $k$ is given by
\begin{equation}
\begin{split}
h_k(t)=F^k_{+}(\mathbf{r}_k,\theta,\phi,\psi)\, h_{+}(t-\delta t_k) \\
      +F^k_{\times}(\mathbf{r}_k,\theta,\phi,\psi)\, h_{\times}(t-\delta t_k),
\end{split}
\end{equation}
where $F^k_{+}$ and $F^k_{\times}$ are the detector antenna response functions for the plus and cross polarizations, respectively, and $\theta,\phi,\psi$ denote the sky location and polarization angle of the source in the Earth-fixed frame. The arrival-time delay between the Earth center and detector $k$ is
\begin{equation}
\delta t_k = \frac{\mathbf{r}_k \cdot \hat{\mathbf{n}}}{c},
\end{equation}
where $\mathbf{r}_k$ is the detector position vector, $\hat{\mathbf{n}}$ is the GW propagation direction, and $c$ is the speed of light.

The observed data stream at detector $k$ is a linear combination of signal and noise,
\begin{equation}
d_k(t) = n_k(t) + h_k(t),
\end{equation}
where $n_k(t)$ denotes the detector noise. 

To account for observational variability, we consider randomly oriented sources. For axisymmetric CCSN simulations, the GW amplitude scales approximately as $\propto \sin^2\alpha$, where $\alpha$ is the angle between the stellar rotation axis and the line of sight \citep{ott12correlated}. Since the emitted GW signal is axisymmetric, the cross polarization vanishes ($h_{\times}=0$), and only the plus polarization contributes to the detector response. 
We inject signals in time domain. Before injection, to suppress edge effects and reduce spectral leakage, each signal is tapered using a Tukey window~\citep{tukey1967intoroduction} with $\alpha=0.1$. The data is then bandpass-filtered between 20 and 2000~Hz and whitened using the detector PSD. 
The processed strain is segmented into 70~ms windows, while the CCSN signals themselves have a typical duration of 8~ms. To account for the uncertainty in the core-bounce time, each signal is injected with a random bounce-time uncertainty of $\Delta t_{\rm b}=20$~ms.

After preprocessing, each strain segment is transformed into the frequency domain using a Fourier transform. This choice is motivated in part by the uncertainty in the core-bounce time, which introduces temporal shifts between signals and can hinder feature learning by a conventional CNN operating on time-domain waveforms. In contrast, a time shift alters the Fourier phase but preserves the amplitude spectrum, making the latter inherently robust to temporal misalignment. This motivation is consistent with our previous study, which demonstrated improved CCSN parameter estimation performance using frequency-domain rather than time-domain inputs~\citep{Akhmetali26PE}. The resulting frequency-domain representation is used as input to the DL model described in Section~\ref{sec:CNN}.

\subsection{Deep learning parameter estimation}
\label{sec:CNN}

DL has become one of the most powerful approaches for extracting complex nonlinear relationships from high-dimensional data~\citep{Lecun15deep}. Among DL methods, Convolutional Neural Networks (CNNs) are particularly effective for analyzing structured signals because they automatically learn hierarchical feature representations directly from the input data through convolutional operations~\citep{Gu18CNN, Akhmetali24, Ussipov24}. 

\begin{table*}[htbp]
\centering
\caption{Architecture of the CNN model used in this study.}
\label{tab:cnn}
\begin{tabular}{lllcc}
\hline
Layer & Type & Parameters & Output Shape & Activation \\
\hline
0   &   Conv1D          &   32 filters, kernel size 3   &   $(142, 32)$& ReLU  \\
1   &   MaxPooling1D    &   Pool size 2                 & $(71, 32)$   &  \\
2   &   Conv1D          &   64 filters, kernel size 3   & $(69, 64)$   & ReLU  \\
3   &   MaxPooling1D    &   Pool size 2                 & $(34, 64)$   &  \\
4   &   Conv1D          &   128 filters, kernel size 3  & $(32, 128)$  & ReLU  \\
5   &   MaxPooling1D    &   Pool size 2                 & $(16, 128)$  &  \\
6   &   Flatten         &                               & $(2048)$     &  \\
7   &   Dense           &   512 units                   & $(512)$      & ReLU \\
8   &   Dense           &   256 units                   & $(256)$      & ReLU \\
9   &   Dense           &   3 units                     & $(3)$        & Linear \\
\hline
\end{tabular}
\end{table*}

In this work, we employ a one-dimensional CNN to estimate the source parameters from the detector network outputs. The use of a CNN is a natural choice because the neural network receives multi-channel inputs containing correlated information from multiple GW detectors.

Table~\ref{tab:cnn} summarizes the architecture adopted in this study. The neural network consists of three convolutional blocks, each composed of a Conv1D layer followed by a MaxPooling1D layer. The convolutional layers use 32, 64, and 128 filters, respectively, with a kernel size of 3 and ReLU activation functions. These layers progressively extract increasingly complex features from the detector data while the pooling operations reduce the dimensionality of the feature maps and improve computational efficiency.

After the final convolutional block, the feature maps are flattened into a one-dimensional feature vector of length 2048. This representation is then passed through two fully connected layers containing 512 and 256 neurons, respectively, each employing ReLU activation. Finally, a linear output layer produces the predicted source parameters ($f_{\mathrm{peak}}$, $T/|W|$, and $D\Delta h$). The complete neural network contains approximately $1.21 \times 10^{6}$ trainable parameters (i.e., the weights and biases optimized during training).

We train the neural network using the Adam optimizer~\citep{kingma17adam} with an initial learning rate of $10^{-3}$. The model parameters are optimized by minimizing the mean absolute error between the predicted and true parameter values. To improve convergence, we employ a learning-rate scheduler that reduces the learning rate by a factor of two whenever the validation loss does not improve for ten consecutive epochs, with a minimum learning rate of $10^{-6}$. In addition, we apply early stopping with a patience of 20 epochs and restore the model weights corresponding to the lowest validation loss to mitigate overfitting and improve generalization. 

To evaluate the generalization performance of the CNN, we randomly reserve 15\% of the original waveform catalogue as an independent test set (200 signals), while the remaining 85\% is used for model development. At each iteration, the development set is randomly divided into training and validation subsets, corresponding to 70\% and 15\% of the original catalogue, respectively. This procedure is repeated ten times, with independently randomized training and validation subsets and freshly initialized CNN models, while the test set remains fixed throughout all iterations. The input features and target parameters are standardized using statistics calculated exclusively from the training subset, and the same transformations are applied to the validation and test subsets\footnote{We use \texttt{StandardScaler} from \texttt{scikit-learn}~\citep{pedregosa2011scikit} for this standardization.}. For each iteration, the model is trained from scratch and evaluated on the same independent test set, which is excluded from all training, validation, and model-selection procedures. The GW injections, noise realizations, and waveform--sky-position assignments remain fixed across iterations, ensuring that variations in the resulting performance metrics reflect differences in model initialization and the selection of training and validation samples. We report the mean and standard deviation of the performance metrics across the ten iterations.

To quantify the performance of our parameter estimation framework, we use the coefficient of determination ($R^2$), defined as
\begin{equation}
R^2 = 1 - \frac{\sum_{i=1}^{N}(y_i-\hat{y}i)^2}
{\sum_{i=1}^{N}(y_i-\bar{y})^2},
\end{equation}
where $y_i$ and $\hat{y}_i$ are the true and predicted values of the target parameter for the $i$th sample, $\bar{y}$ is the mean of the true values, and $N$ is the total number of samples. The $R^2$ statistic measures the fraction of the variance in the target parameter that is explained by the model predictions. A value of $R^2=1$ indicates perfect agreement between predictions and observations, whereas $R^2=0$ corresponds to a model that performs no better than predicting the mean of the target variable. Negative values indicate performance worse than this baseline. Consequently, larger $R^2$ values correspond to more accurate parameter recovery, with values approaching unity indicating excellent predictive performance~\citep{coulibaly_r2}.

\section{Results}
\label{sec:results}

To assess the performance of the parameter estimation framework, we use the CCSN waveforms and source locations presented in Sections~\ref{sec:data} and~\ref{sec:CCSN}, respectively, under different detector network configurations and distances. In all cases, we estimate the peak frequency $f_{\mathrm{peak}}$, rotation rate $T/|W|$, and the peak amplitude $D\Delta h$, and compare the predictions with their true values. In our analysis below, we fix the GPS time to $t_{\rm GPS}=1325048418$. This time is chosen arbitrarily and its value does not affect our conclusions. 

\subsection{Parameter estimation distance}
\label{sec:PE}

We first assess the distance reach of our parameter estimation framework using current-generation (2G) detector networks. As the distance increases, the signal-to-noise ratio decreases, making parameter recovery more challenging. Consequently, the regression performance is strongly determined by the detector network sensitivity.

In this experiment, we compare how the parameter estimation performance ($R^2$) evolves with source distance for two configurations: the two LIGO interferometers (HL) and the full LIGO--Virgo--KAGRA network (HLVKA). We vary the source distance. 

\begin{figure}[t]
\centering
\includegraphics[width=1\linewidth]{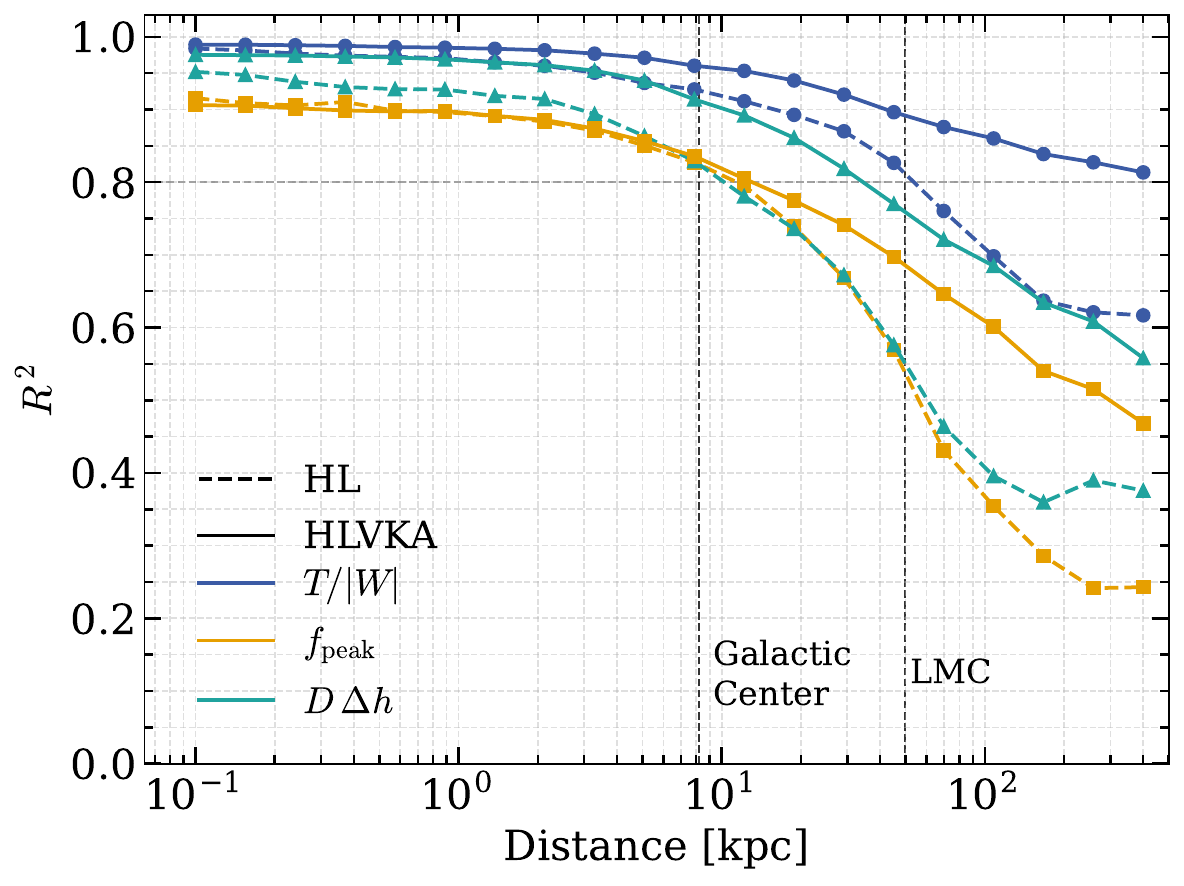}
    \caption{Mean $R^2$ score as a function of distance for different parameters and detector networks. The blue, orange and green lines correspond to $T/|W|$, $f_{\mathrm{peak}}$, and $D\Delta h$, respectively. Dashed lines represent HL detector network, while solid lines correspond to HLVKA detector network. The gray horizontal dashed line at $R^2=0.8$ represents a limit for reasonably accurate estimations. Vertical black dashed lines mark reference distances to the Galactic Center (8.2~kpc), and the Large Magellanic Cloud (49.8~kpc).} 
     \label{fig:distance}
\end{figure}

Fig.~\ref{fig:distance} shows the evolution of the mean $R^2$ score as a function of source distance for the HL and HLVKA detector networks. As expected, the parameter-estimation accuracy decreases with increasing distance as the GW signals become progressively weaker relative to the detector noise. The HLVKA network consistently achieves higher $R^2$ scores than HL, with the difference becoming more pronounced at larger distances. However, the extent of this improvement depends on the parameter being estimated.

Adopting $R^2=0.8$ as the threshold for accurate parameter estimation, the distance reach for the $f_{\mathrm{peak}}$ is approximately $10$--$15$~kpc for both networks. For $D\Delta h$, the distance reach increases from $\sim10$~kpc with HL to $\sim35$~kpc with HLVKA. A substantially greater improvement is observed for $T/|W|$, whose distance reach extends from $\sim50$~kpc with HL to beyond $100$~kpc with HLVKA. Among the three parameters, $T/|W|$ therefore exhibits the greatest robustness to increasing source distance, particularly when using the larger detector network.

Unlike \citet{Bruel23Inference}, who distinguish between ``favourable'' and ``unfavourable'' detector configurations using the equivalent antenna pattern $F_{\rm eq}$, we fix the GW arrival time $t_{\rm GPS}$ and adopt the Galactic CCSN angular distribution described in Section~\ref{sec:CCSN}. Therefore, our distance-reach estimates characterize the mean parameter estimation performance for the adopted Galactic source population rather than particular favourable or unfavourable sky configurations.

Overall, the current detector network enables accurate recovery of all three parameters for rotating Galactic CCSNe, with the potential of reaching further distances with larger detector networks.

\subsection{Sky coverage}
\label{sec:sky_cov}

Having established the distance reach of the parameter estimation framework, we now investigate its sky coverage. Since the detector response depends on the source position through the antenna pattern functions, the estimation accuracy is expected to vary across the celestial sphere. The addition of detectors with different orientations and geographical locations can therefore improve the parameter estimation performance across different sky directions.

In this experiment, we fix the source distance at 8.2~kpc and adopt the Galactic CCSN angular distribution described in Section~\ref{sec:CCSN}. We consider four detector-network configurations: HL, HLV, HLVK, and HLVKA. Each waveform is assigned a fixed sky position, which remains unchanged across all network configurations and training iterations.

\begin{figure*}[htbp]
    \centering
    \includegraphics[width=0.95\linewidth]{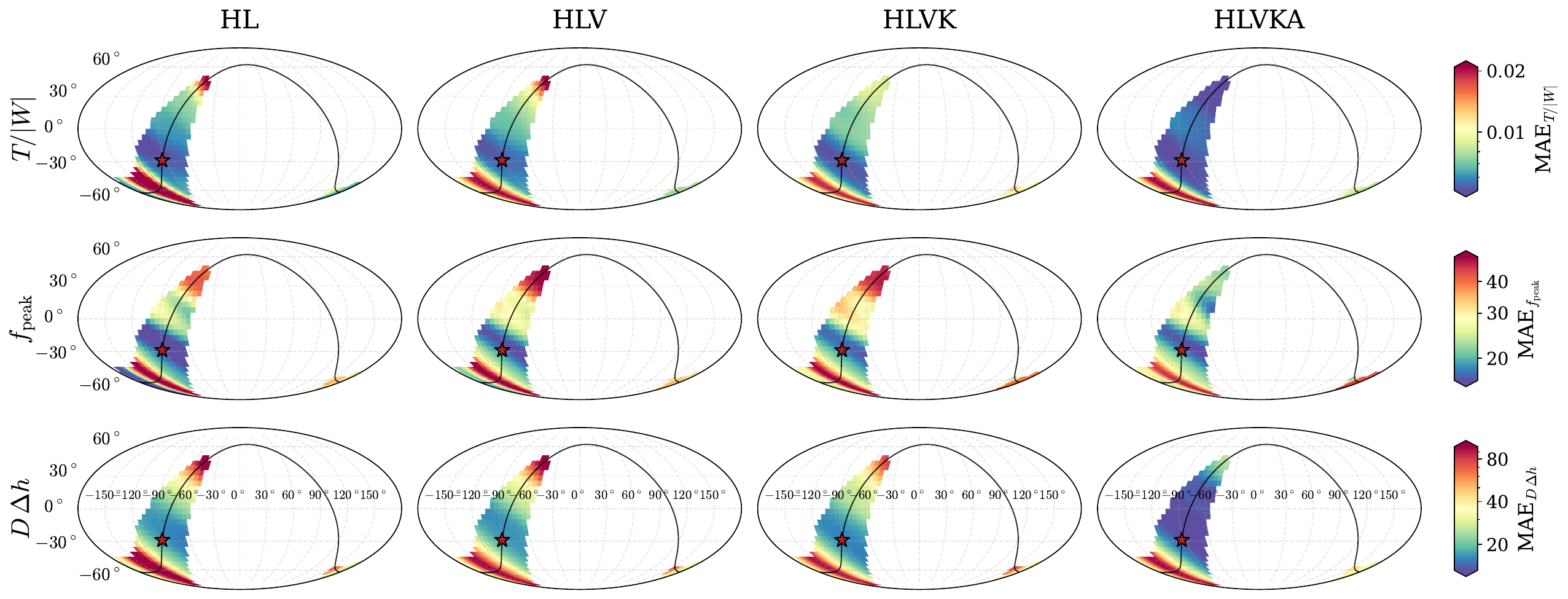}
    \caption{Spatial distribution of the mean absolute error (MAE) for the three estimated parameters and four detector-network configurations at a fixed source distance of 8.2~kpc. Columns correspond to HL, HLV, HLVK, and HLVKA, while rows represent $T/|W|$, $f_{\mathrm{peak}}$, and $D\Delta h$. Colors indicate the MAE on separate logarithmic scales for each parameter, with warmer colors corresponding to larger errors. The black curve represents the Galactic plane, and the red star indicates the Galactic center. Regions with insufficient source sampling are left blank.}
    \label{fig:skymap}
\end{figure*}

To quantify the spatial variation in parameter estimation accuracy, we calculate the absolute prediction error for each waveform in the held-out test set and average it over the iterations in which the waveform is tested. The resulting errors are spatially smoothed using a Gaussian kernel, while regions with insufficient sampling are excluded. Figure~\ref{fig:skymap} presents the resulting mean absolute error (MAE) distributions for $T/|W|$, $f_{\mathrm{peak}}$, and $D\Delta h$ in equatorial coordinates. Separate logarithmic color scales are used to account for the different magnitudes of the three parameters.

As shown in Fig.~\ref{fig:skymap}, all three parameters exhibit broadly similar spatial distributions of MAE, with relatively low errors around the Galactic center and higher errors toward positive declinations and the southernmost regions of the sampled sky. This spatial dependence is consistent with variations in the directional sensitivity of the detector networks. However, since each sky position is associated with a different waveform, differences in intrinsic waveform properties may also contribute to the observed variations.

The addition of detectors generally reduces the MAE, although the magnitude of the improvement depends on the network configuration. In particular, adding KAGRA to HLV noticeably improves the estimation of $T/|W|$ in the region approximately bounded by RA$=-60^\circ$ to $-30^\circ$ and Dec$=30^\circ$ to $60^\circ$. The greatest improvement is observed when LIGO Aundha is included in the HLVKA network, which exhibits considerably lower errors across much of the sampled sky for all three parameters. This can be partly attributed to the assumed noise sensitivity of LIGO Aundha, which is comparable to that of LIGO Hanford and Livingston and generally higher than that of Virgo and KAGRA over the relevant frequency range (see Fig.~\ref{fig:ASDs}). Combined with its distinct geographical location and orientation, this additional sensitivity improves the overall parameter estimation accuracy and provides more uniform coverage of the sampled Galactic source directions.

\subsection{Third-generation detectors}
\label{sec:third_gen}

Finally, we assess the distance reach of our parameter estimation framework using 3G detectors. As discussed in Section~\ref{sec:detectors}, the substantially improved sensitivity of these observatories is expected to extend the distance over which CCSN parameters can be accurately estimated.

In this experiment, we repeat the analysis described in Section~\ref{sec:PE} using the detector network consisting of three detectors (ET-CE20-CE40). The source distance is varied from 0.1~kpc to 3~Mpc, while the waveform--sky-position assignments follow the same Galactic CCSN angular distribution adopted in the previous experiments. 

\begin{figure}[t]
    \centering
    \includegraphics[width=\linewidth]{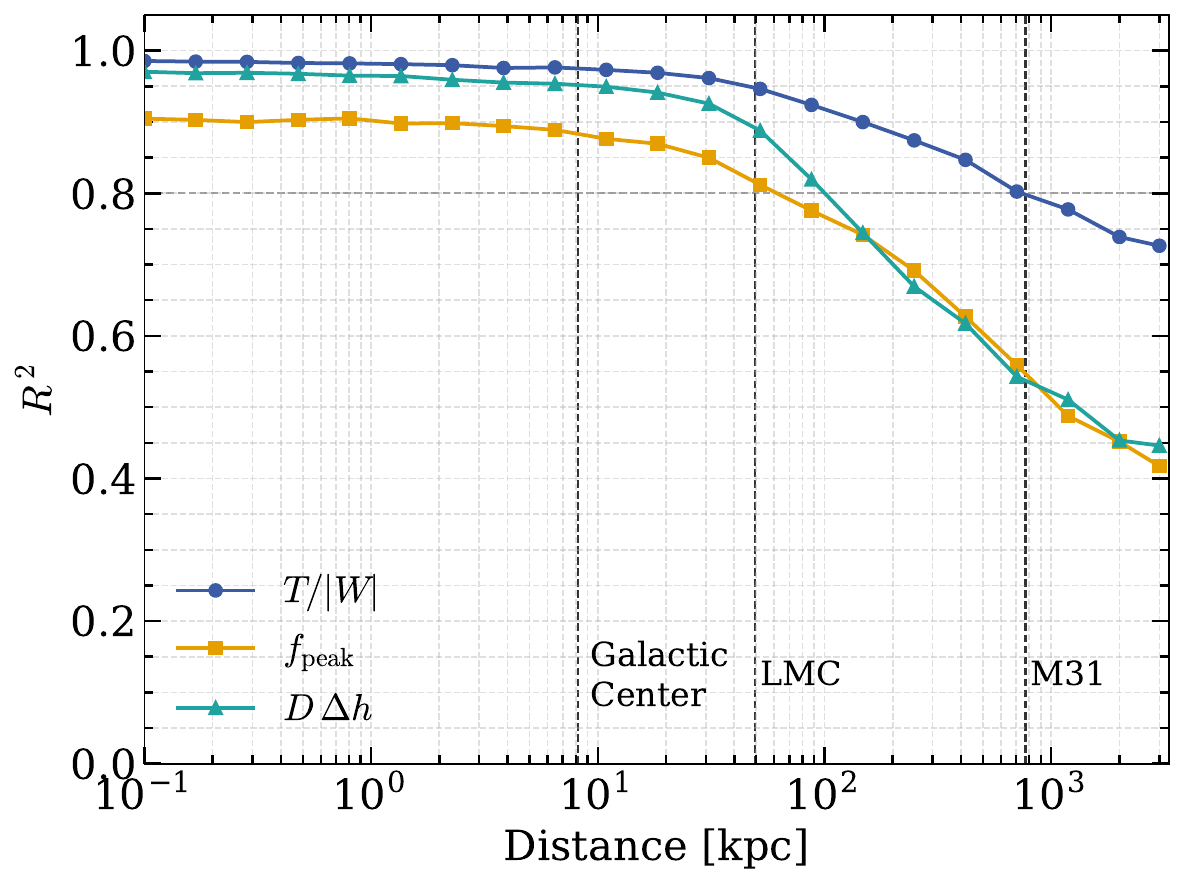}
    \caption{Mean $R^2$ score as a function of distance for different parameters with ET-CE20-CE40 detector network. The blue, orange and green lines correspond to $T/|W|$, $f_{\mathrm{peak}}$, and $D\Delta h$, respectively. The gray horizontal dashed line at 0.8 represents a limit for reasonably accurate estimations. Vertical black dashed lines indicate distances to the Galactic Center (8.2~kpc), the Large Magellanic Cloud (49.8~kpc), and Andromeda (780 kpc).}
    \label{fig:3G}
\end{figure}

Fig.~\ref{fig:3G} shows the evolution of the $R^2$ score as a function of source distance. Compared to the current-generation detector network (cf.~Fig.~\ref{fig:distance}), the 3G configuration substantially extends the distance reach for all three parameters. $f_{\mathrm{peak}}$ remains recoverable out to $\sim60$~kpc, while the corresponding reach for $D\Delta h$ extends to $\sim100$~kpc. A significantly greater distance reach is obtained for $T/|W|$, which maintains $R^2\gtrsim0.8$ out to $\sim800$~kpc.

At the distance of the Large Magellanic Cloud ($49.8$~kpc), all three parameters can be estimated with relatively high accuracy, with $R^2\gtrsim0.8$. However, the differences among the parameters become increasingly pronounced at larger distances. In particular, at the distance of the Andromeda galaxy ($770$~kpc), $T/|W|$ retains an $R^2$ score of $\sim0.8$, whereas the scores for $f_{\mathrm{peak}}$ and $D\Delta h$ decrease to $\sim0.55$ and $\sim0.54$, respectively. These results indicate that $T/|W|$ is substantially more robust to decreasing signal strength than the other two parameters, allowing its accurate estimation at distances approaching the Local Group scale.

\section{Conclusion}
\label{sec:conclusion}

In this work, we have presented a DL framework for estimating the physical properties of rotating CCSNe from their GW signals. Building upon our previous single-detector study~\citep{Akhmetali26PE}, we have extended the methodology to detector networks consisting of current- and third-generation GW observatories. Using CNNs, we have estimated the peak frequency, $f_{\mathrm{peak}}$, rotation rate, $T/|W|$, and peak amplitude, $D\Delta h$ directly from noisy detector data under realistic observing conditions. We have adopted a Galactic population model based on an exponential disk, allowing us to assess parameter-estimation performance across a physically motivated distribution of source directions. Our study has been  further motivated from the work of \cite{Bruel23Inference} who showed that a network of 3G detectors can significantly improve the recovery of PNS properties for nonrotating CCSN models. 

Our results demonstrate that the parameter-estimation accuracy depends strongly on both the source distance and the detector-network configuration. For 2G detectors, the distance reach for $f_{\mathrm{peak}}$ is $\sim10$--$15$~kpc. The corresponding reach for $D\Delta h$ increases from $\sim10$~kpc with HL to $\sim35$~kpc with HLVKA. A more substantial improvement is obtained for $T/|W|$, whose distance reach extends from $\sim50$~kpc with HL to beyond $\sim100$~kpc with HLVKA. Among the three parameters, $T/|W|$ exhibits the greatest robustness to increasing source distance, while the recovery of $f_{\mathrm{peak}}$ remains the most challenging (cf.~Section~\ref{sec:PE}).

We have also investigated the spatial dependence of parameter-estimation accuracy using the adopted Galactic CCSN population. The resulting sky maps reveal broadly similar regions of relatively high and low accuracy for all three parameters. The addition of detectors generally reduces the estimation errors across the sampled sky, with the most substantial improvement observed when LIGO Aundha is included in the HLVKA network. This improvement can be partly attributed to its assumed sensitivity, comparable to that of LIGO Hanford and Livingston, together with its distinct geographical location and orientation. These findings highlight the importance of both detector sensitivity and network geometry for improving parameter-estimation performance across Galactic source directions (cf.~Section~\ref{sec:sky_cov}).

Finally, we have assessed the capabilities of 3G detector networks consisting of the ET and CE observatories. Owing to their substantially improved sensitivities, accurate estimation of $f_{\mathrm{peak}}$ and $D\Delta h$ becomes possible out to $\sim60$~kpc and $\sim100$~kpc, respectively. The greatest distance reach is achieved for $T/|W|$, which remains accurately recoverable out to $\sim800$~kpc. At the distance of Andromeda ($770$~kpc), the estimation of $T/|W|$ retains an $R^2$ score of approximately $0.8$, demonstrating the potential of 3G detector networks to constrain the rotational properties of rapidly rotating CCSNe beyond the Milky Way (cf. Section~\ref{sec:third_gen}).

Compared to the results of \citet{Bruel23Inference}, we find larger distance reaches for parameter estimation. While their study considered non-rotating CCSNe, our analysis focuses on rotating progenitors and incorporates a physically motivated Galactic CCSN angular distribution. Although differences in the adopted source populations and methodologies should be taken into account, our results suggest that the physical properties of rotating CCSNe may be recoverable at greater distances than those of non-rotating events. The most likely reason is that rotating CCSNe produce stronger GW emission during core bounce and the early post-bounce phase, resulting in higher signal-to-noise ratios and more robust parameter recovery~\citep{Szczepanczyk21Detecting}.

Although these results demonstrate the strong potential of DL for CCSN parameter estimation, several limitations remain. First, our analysis relies on 2D axisymmetric simulations of rotating CCSNe, which provide only the plus polarization of the GW signal. However, previous studies have shown that 2D and 3D simulations produce similar GW signals during core bounce, as rotating stellar cores remain largely axisymmetric at this stage~\citep{ott:07cqg, ott12correlated}. Three-dimensional effects become important during the subsequent post-bounce evolution. Therefore, we expect their influence to be relatively limited for the bounce-dominated signals considered here, although a quantitative assessment requires further investigation using 3D simulations.

Furthermore, our analysis considers only the bounce and early post-bounce GW signal, which does not capture the full complexity of CCSN emission, and assumes stationary Gaussian detector noise. Although the training, validation, and test subsets contain distinct waveforms, related simulations with similar physical parameters may appear in different subsets. Therefore, the reported performance reflects the model's ability to generalize to unseen waveforms from the same simulation catalogue rather than entirely independent simulation models. Addressing these limitations, particularly through the inclusion of 3D simulations and more realistic detector noise, will be the focus of future work.

\begin{acknowledgments}
We thank António Onofre, Nurzhan Ussipov and Marat Zaidyn for fruitful discussions during the course of this work. We also thank Nelson Christensen for his comments on this manuscript. We are grateful to the anonymous referee for their careful reading of the manuscript and constructive comments and suggestions, which helped improve the quality and clarity of this work.

This research was funded by the Science Committee of the Ministry of Science and Higher Education of the Republic of Kazakhstan (Grant No. AP26103591). EA is partially supported by the Nazarbayev University Faculty Development Competitive Research Grant Program (no. 040225FD4713). 
SN was supported by FCT - Fundação para a Ciência e Tecnologia, I.P. by project reference 2025.02005.BD and DOI identifier https://doi.org/10.54499/2025.02005.BD. 
SN acknowledges financial support by FCT in the framework of the Strategic Funding UID/04650/2025.
JAF is supported by the Spanish Agencia Estatal de Investigación (grant PID2024-159689NB-C21) funded by MICIU/AEI/10.13039/501100011033 and by FEDER / EU, and by the Generalitat Valenciana (Prometeo Excellence Programme grant CIPROM/2022/49). 

AI-based language tools were used to improve the clarity and grammar of the manuscript. 
\end{acknowledgments}

\appendix

\section{Effect of CNN architecture}
\label{sec:appendix_a}

To assess the influence of the CNN architecture on parameter-estimation performance, we investigate several modifications to the baseline model presented in Table~\ref{tab:cnn}. In particular, we vary the number of convolutional layers, the number of filters, the convolutional kernel size, and the number of fully connected layers. Each modification is considered separately, while the remaining architectural parameters are kept unchanged.

For this comparison, we adopt the HLVKA detector network and fix the source distance at 8.2~kpc. We use the 85\% development set described in Section~\ref{sec:CNN}, excluding the independent test set reserved for the final evaluation. The development set is further divided into training, validation, and internal test subsets containing 70\%, 15\%, and 15\% of the development waveforms, respectively. We repeat the analysis five times using independently randomized data partitions and freshly initialized models. All configurations are trained using the same optimization procedure described in Section~\ref{sec:CNN}.

\begin{figure}[t]
    \centering
    \includegraphics[width=\linewidth]{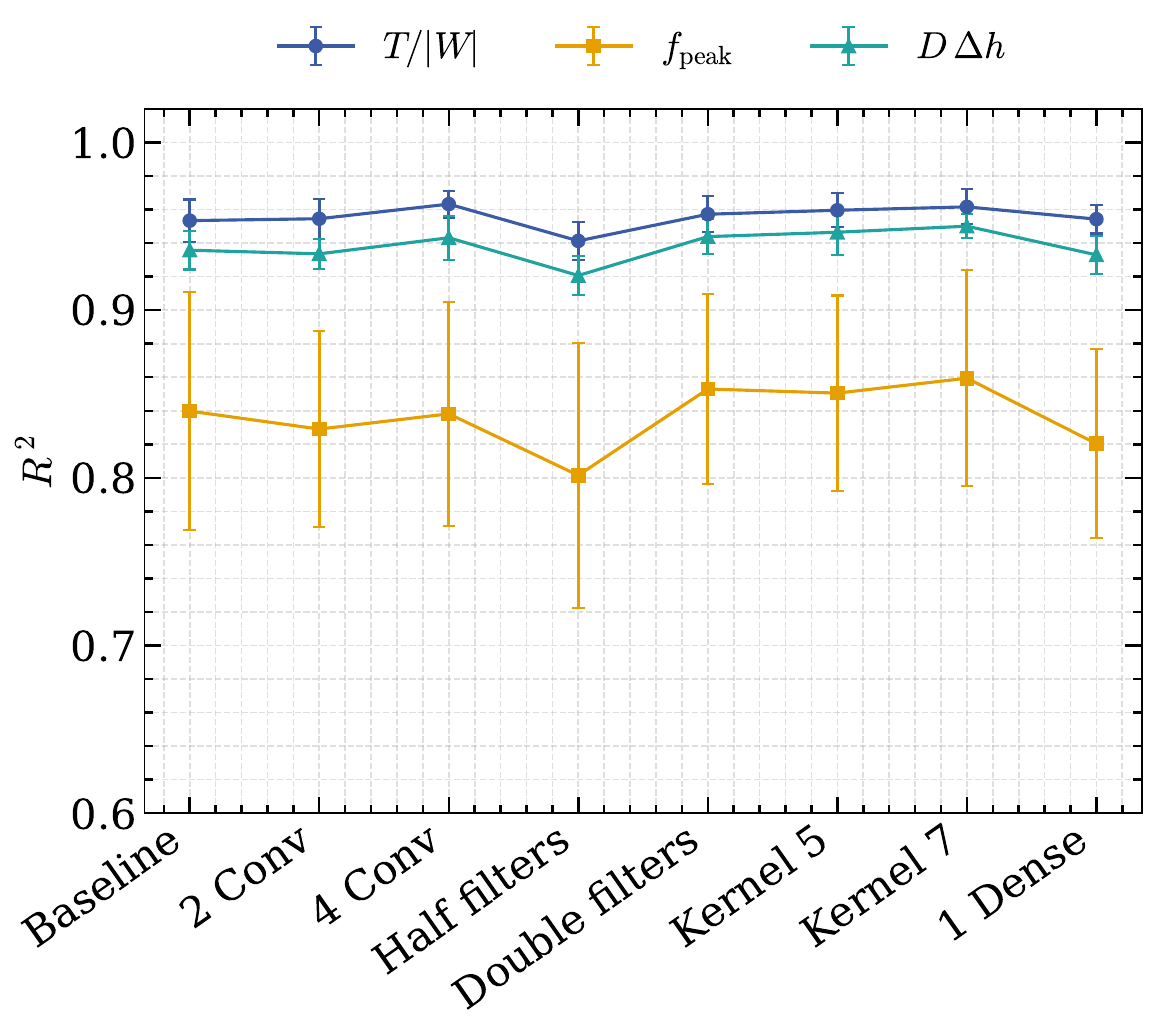}
    \caption{Mean $R^2$ scores for different CNN architectures using the HLVKA detector network at a fixed source distance of 8.2~kpc. The blue, orange, and green lines correspond to $T/|W|$, $f_{\mathrm{peak}}$, and $D\Delta h$, respectively. The error bars represent the standard deviation across five independent training iterations. All configurations are evaluated using the development set, without accessing the independent test set reserved for the final evaluation.}
    \label{fig:CNN_architecture}
\end{figure}

Fig.~\ref{fig:CNN_architecture} shows the mean $R^2$ scores obtained for the three estimated parameters. The baseline configuration consists of three convolutional layers with 32, 64, and 128 filters, a kernel size of 3, and two fully connected layers containing 512 and 256 neurons. We compare this architecture with configurations containing two or four convolutional layers, half or twice the number of filters, kernel sizes of 5 and 7, and a single fully connected layer.

The results indicate that the parameter-estimation performance is relatively insensitive to moderate changes in the CNN architecture. The highest $R^2$ scores are consistently obtained for $T/|W|$, followed by $D\Delta h$, while the estimation of $f_{\mathrm{peak}}$ remains the most challenging. Reducing the number of filters generally decreases the performance for all three parameters, particularly for $f_{\mathrm{peak}}$. In contrast, increasing the number of convolutional layers or the kernel size leads to modest improvements in some cases. However, these improvements are generally comparable to the variations between training iterations.

Overall, the baseline architecture provides competitive performance across all three parameters without requiring additional convolutional layers or larger kernels. We therefore retain this configuration for the parameter-estimation analysis presented in the main text.

\section{Comparison with other models}
\label{sec:appendix_b}

To further assess the suitability of the adopted CNN for parameter estimation, we compare its performance with two alternative DL architectures: a fully connected neural network (FCNN) and a one-dimensional residual neural network (ResNet1D). The FCNN consists of two fully connected layers with 512 and 256 neurons, followed by a linear output layer. The ResNet1D contains three residual convolutional blocks with 32, 64, and 128 filters, respectively. Each block employs convolutional layers with a kernel size of 3 and a shortcut connection, followed by max-pooling. The resulting features are passed through two fully connected layers containing 512 and 256 neurons. All three models predict the same physical parameters, $T/|W|$, $f_{\mathrm{peak}}$, and $D\Delta h$.

We perform the comparison using the HLVKA detector network at a fixed source distance of 8.2~kpc. As in Section~\ref{sec:appendix_a}, only the 85\% development set is used, with an internal training, validation, and test split of 70\%, 15\%, and 15\%, respectively. The analysis is repeated five times, and the same data partitions, input representations, and optimization settings are used for all three models to ensure a consistent comparison.

\begin{figure}[t]
    \centering
    \includegraphics[width=\linewidth]{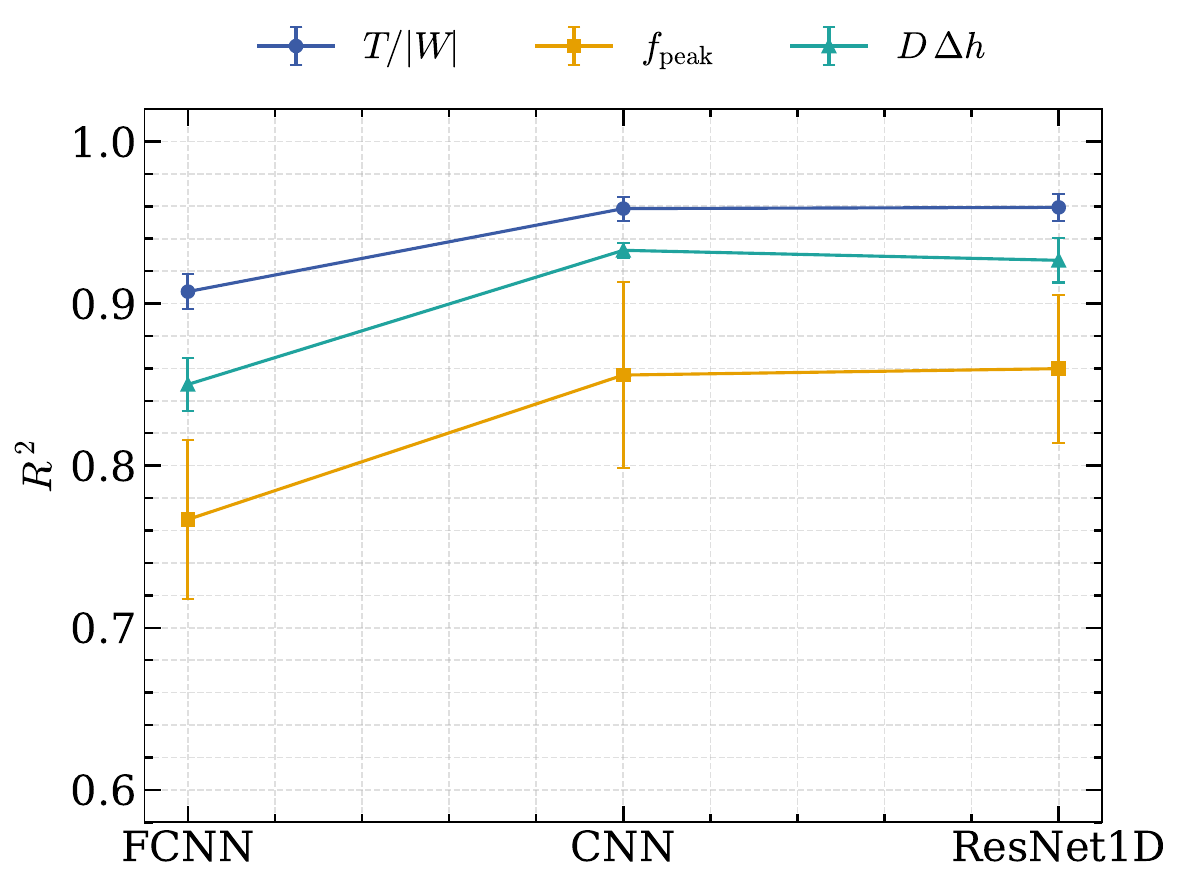}
    \caption{Comparison of the mean $R^2$ scores obtained with the FCNN, CNN, and ResNet1D models using the HLVKA detector network at a source distance of 8.2~kpc. The blue, orange, and green lines correspond to $T/|W|$, $f_{\mathrm{peak}}$, and $D\Delta h$, respectively. The results are evaluated on the internal test subset of the development data, and the error bars indicate the standard deviation across five independent training iterations.}
    \label{fig:model_comparison}
\end{figure}

Fig.~\ref{fig:model_comparison} compares the parameter-estimation performance of the three architectures. The FCNN generally exhibits the lowest $R^2$ scores, with values of approximately 0.91, 0.77, and 0.85 for $T/|W|$, $f_{\mathrm{peak}}$, and $D\Delta h$, respectively. Introducing convolutional layers substantially improves the results, with the CNN achieving corresponding scores of approximately 0.96, 0.86, and 0.93. This improvement suggests that convolutional operations are effective in extracting relevant features from the frequency-domain GW signals.

The ResNet1D achieves performance comparable to that of the CNN, with only minor differences between the two models. In particular, both architectures provide nearly identical accuracy for $T/|W|$ and $f_{\mathrm{peak}}$, while the CNN performs slightly better for $D\Delta h$. These results indicate that the additional residual connections do not provide a substantial improvement for the present dataset and parameter-estimation task. Consequently, the adopted CNN offers a suitable compromise between architectural complexity and predictive performance, supporting its use throughout the main analysis.

\bibliography{gw_sn}

\end{document}